\documentclass{article}
\usepackage[utf8]{inputenc}
\usepackage{amsmath}
\usepackage[pdftex]{graphicx}
\usepackage{float}
\usepackage{amscd}
\usepackage{calrsfs}
\usepackage{slashed}
\usepackage{amsfonts}
\usepackage{authblk}
\usepackage[letterpaper, portrait, margin=0.7in]{geometry}
\title{\bf Principal equatorial null geodesic congruences in the Kerr metric, and their quantum propagators}
\author{Josu\'e G. Mateos Trujillo$^{1} $ and Miguel Socolovsky$^{1,2}$}
\affil{$^{1}$Instituto de Ciencias Nucleares, Universidad Nacional Aut\'onoma de M\'exico, Cd. Universitaria, 04510, Ciudad de M\'exico, M\'exico\\$^{2}$Instituto de Astronom\'\i a y F\'\i sica del Espacio, Universidad de Buenos Aires-CONICET, Argentina\\socolovs@nucleares.unam.mx}

\providecommand{\keywords}[1]
{
	\small	
	\textbf{Keywords:} #1
}
\begin{document}
\date{}
\maketitle

\begin{abstract}
Using the Raychaudhuri equation, we associate quantum probability amplitudes (propagators) to equatorial principal ingoing and outgoing null geodesic congruences in the Kerr metric. The expansion scalars diverge at the ring singularity; however, the propagators remain finite, which is an indication that at the quantum level singularities might disappear or, at least, become softened.	
\end{abstract}
\keywords{Kerr metric, principal null geodesics, propagators}

\section{Introduction}

\

In a recent paper [1], Feynman propagators were associated to timelike and null geodesic congruences in the Schwarzschild spacetime. The possibility of this association lies in the Raychaudhuri equation [2] which describes the flow of such congruences, even in the case of curves where an acceleration term is present. In particular, the equation for the expansion scalar $\Theta$, after a transformation of variables is done [3], becomes a 1-dimensional free harmonic oscillator equation for a function $F$ with a  ``time" dependent frequancy (the rôle of time is played by the affine parameter $\lambda$ of the geodesics). A Lagrangian can be immediately constructed, and with initial and final values for $\lambda$ and for the function $F$ (basically representing the transverse area of the congruences), an exact path integral $K(F'',\lambda'';F',\lambda')$ of the exponential of the corresponding action is obtained. The basic idea is that this $K$ describes the quantum evolution of the congruences, even in the absence, at present, of a final theory of quantum gravity [4]. 

\

In the present paper, we restrict the analysis to the simplest case in the Kerr spacetime [5]: the forward and past directed, ingoing and outgoing, principal null geodesics in the equatorial plane ($\theta=\pi/2$), and their congruences. Since the Kerr's solution in the Petrov's classification [6] is algebraically special, the Goldberg-Sachs theorem [7] implies that the shear $\sigma_{\mu\nu}$ in the Raychaudhuri equation vanishes and, 
since an explicit calculation shows that the rotation term $\omega_{\mu\nu}$ also vanishes, the above mentioned Lagrangian, as in the Schwarzschild case, reduces to that of a free non relativistic particle. At the ring singularity ($r=0,\ \theta=\pi/2$) the expansion diverges as $1/r$. However, the relevant propagators $K$ remain finite. This result, though does not constitute a proof that singularities disappear in a quantum treatment of black holes, is however an indication that at the quantum level they might indeed disappear or, at least, become smoother. 

\

In section 2, following the presentation of Ferrari et al [8], we integrate the equations of motion of principal equatorial null geodesics {\it p.e.n.g.}'s, identify the tangent vector field $k^\mu$ for them, and consequently determine the affine parameter $\lambda$ as $r$ for outgoing geodesics and $-r$ for ingoing ones. In the process one proves that the Carter constant [9]  reflecting a hidden symmetry of the Kerr solution, vanishes for {\it p.e.n.g.}'s. In Figures 1 and 2 we plot $t(r)$ and $\psi(r)$ for forward and past directed geodesics where $t$, $r$, and $\psi$ are the usual Boyer-Lindquist coordinates ($    
\theta=\pi/2$ is fixed at the equator). It is interesting to notice that in the ``shell horizon" [10], that is, the region between the event (outer) and Cauchy (inner) horizons, precisely the only region where trapped surfaces exist, all {\it p.e.n.g.}'s, both outgoing and ingoing, are past directed. In all cases, the geodesics are asymptotic to at least one horizon.\

In section 3 we construct the Raychaudhuri equation that adquires the simplest form as in the Schwarzschild case due to the vanishing of both the shear and the rotation terms (the second one because the tensor $B_{\mu\nu}=k_{\mu;\nu}$ turns out to be symmetric). 

\

In section 4 we calculate the potentially divergent propagators associated with geodesic congruences in the region $-\infty<r<r_-$ where the ring singularity lies, and prove their finitness. Finally, section 5 is deserved to conclusions and remarks.

\

We use the natural system of units: $G=c=\hbar=1$ and the metric signature $(+,-,-,-)$.

\

\section{Principal equatorial null geodesics}

\

In Boyer-Lindquist coordinates $x^\mu=(x^0,x^1,x^2,x^3)=(t,r,\theta,\psi)$ the stationary axial symmetric Kerr metric is given by 
\begin{equation}
ds^2={{\Delta-a^2sin^2\theta}\over{\Sigma}}dt^2-{{\Sigma}\over{\Delta}}dr^2-\Sigma d\theta^2-{{sin^2\theta}\over{\Sigma}}Ad\psi^2+{{2asin^2\theta}\over{\Sigma}}(r^2+a^2-\Delta)dtd\psi,
\end{equation}
where $\Delta=r^2+a^2-2Mr$, $\Sigma=r^2+a^2cos^2\theta$, $A=(r^2+a^2)^2-a^2sin^2\theta \Delta$, with $a=J/M<M$ where $M$ and $J$ are the total energy and total angular momentum parameters of the solution. Also, $\Delta=(r-r_+)(r-r_-)$ with $r_+=M+\sqrt{M^2-a^2}$ ($r_-=M-\sqrt{M^2-a^2}$) the event or exterior horizon $h_+$ (the Cauchy or interior horizon $h_-$). Respectively corresponding to the stationarity and axial symmetries one has the Killing vectors $\partial_t$ and $\partial_\psi$. 

\

The geodesic motion of particles in the metric (1) is described by the Lagrangian
\begin{equation}
{\cal L}={{1}\over{2}}g_{\mu\nu}\dot{x}^\mu\dot{x}^\nu={{1}\over{2}}({{\Delta-a^2sin^2\theta}\over{\Sigma}}\dot{t}^2-{{\Sigma}\over{\Delta}}\dot{r}^2-\Sigma\dot{\theta}^2-{{sin^2\theta}\over{\Sigma}}A\dot{\psi}^2+{{2asin^2\theta}\over{\Sigma}}(r^2+a^2-\Delta)\dot{t}\dot{\psi}),
\end{equation}
where $\dot{x}^\mu={{dx^\mu}\over{d\lambda}}$, with $\lambda=\tau$: proper time for massive ($m$) particles, in which case ${\cal L}=m^2$, and $\lambda$ an arbitrary (up to an affine transformation) affine parameter in the massless case, in which case ${\cal L}=0$. The two conserved quantities associated to the above mentioned symmetries are obtained from the Lagrange equation 
\begin{equation}                                               
{{d}\over{d\lambda}}({{\partial {\cal L}}\over{\partial \dot{x}^\mu}})={{\partial {\cal L}}\over{\partial x^\mu}},
\end{equation}
where 
\begin{equation}
p_\mu={{\partial {\cal L}}\over{\partial\dot{x}^\mu }}=g_{\mu\nu}\dot{x}^\nu.
\end{equation}
For $\mu=0,3$ the r.h.s. of (3) vanishes and one obtains
\begin{equation}
p_0=E=({{1-2Mr}\over{\Sigma}})\dot{t}+{{2Marsin^2\theta}\over{\Sigma}}\dot{\psi}
\end{equation}
and
\begin{equation}
p_3=L=-{{2Mrasin^2\theta}\over{\Sigma}}\dot{t}+{{sin^2\theta}\over{\Sigma}}((r^2+a^2)^2-a^2sin^2\theta\Delta)\dot{\psi}^2,
\end{equation}
for the energy and the angular momentum along the rotation axis. From (5) and (6) one obtains 
\begin{equation}
\dot{t}=\dot{t}(r,\theta;M,a;E,L)={{CE-BL}\over{AC+B^2}} \ \ and \ \ \dot{\psi}=\dot{\psi}(r,\theta;M,a;E,L)={{BE+AL}\over{AC+B^2}},
\end{equation}
with
\begin{equation}
A=g_{tt}=1-{{2Mr}\over{\Sigma}}, \ \ B=g_{t\psi}={{2Mra}\over{\Sigma}}sin^2\theta, \ \ C=-g_{\psi\psi}=(r^2+a^2+{{2Mra^2}\over{\Sigma}})sin^2\theta,
\end{equation}
which can be integrated once the equation for $\dot{r}$ and $\dot{\theta}$ are also found. To this aim, one appeals to the Hamilton-Jacobi equation [9]. The Hamiltonian is 
\begin{equation}
{\cal H}(x^\mu,p_\nu)=p_\mu \dot{x}^\mu-{{1}\over{2}}g_{\mu\nu}
\dot{x}^\mu\dot{x}^\nu={{1}\over{2}}g^{\mu\nu}p_\mu p_\nu=p^2/2=\kappa/2,
\end{equation}  
where $\kappa=m^2$ for massive particles and $\kappa=0$ in the massless case. The Hamilton principal function $S$ is defined by
\begin{equation}
{\cal H}(x^\mu,{{\partial S}\over{\partial x^\nu}})+{{\partial S}\over{\partial \lambda}}=0,
\end{equation} 
where ${{\partial S}\over{\partial x^\nu}}=p_\nu$. Writing 
\begin{equation}
S=-{{1}\over{2}}\kappa\lambda+Et+L\psi+\tilde{S}(r,\theta)
\end{equation}
one looks for a separable solution 
\begin{equation}
\tilde{S}(r,\theta)=\tilde{S}_1(r)+\tilde{S}_2(\theta).
\end{equation}
A long but straightforward procedure [8] leads to the equality 
\begin{equation}
\Delta({{d\tilde{S}_1(r)}\over{dr}})^2+\kappa r^2-{{r^2+a^2}\over{\Delta}}E^2-{{4Mra}\over{\Delta}}EL-{{a^2}\over{\Delta}}L^2+a^2E^2+L^2=-({{d\tilde{S}_2(\theta)}\over{d\theta}})^2+(\kappa+E^2)a^2cos^2\theta-{{cos^2\theta}\over{sen^2\theta}}L^2
\end{equation} 
where the l.h.s. (r.h.s.) depends only on $r$ ($\theta$). So, both sides of (13) depend on a constant: -${\cal C}$, the Carter constant. ($[{\cal C}]=[lenght]^4$.) The existence of this constant of motion is due to a hidden symmetry of the Kerr metric, described by a covariant Killing 2-tensor [11]. So one has the set of four constants
\begin{equation}
\{E,L,\kappa,{\cal C}\}
\end{equation}
in terms of which the geodesic motion can be solved. Defining the functions 
\begin{equation}
T(\theta)={\cal C}+cos^2\theta((\kappa+E^2)a^2-{{1}\over{sin^2\theta}}L^2) \ \ and \ \ R(r)=\Delta(\kappa r^2-(L-aE)^2-{\cal C})+(E(r^2+a^2)-La)^2
\end{equation}
one obtains
\begin{equation}
({{d\tilde{S}_2(\theta)}\over{d\theta}})^2=T(\theta) \ \ and \ \ ({{d\tilde{S}_1(r)}\over{dr}})^2={{R(r)}\over{\Delta^2}};
\end{equation}
therefore 
\begin{equation}
S(t,\psi,r,\theta,\lambda;{\cal C},\kappa,E,L)=-{{1}\over{2}}\kappa\lambda+Et+L\psi+\int^r dr^\prime {{\sqrt{R(r^\prime)}}\over{\Delta(r^\prime)}}+\int^\theta d\theta^\prime\sqrt{T(\theta^\prime)}.
\end{equation}

\

For {\it null geodesics} ({\it n.g.}), $\kappa=0$ and so
\begin{equation}
 R(r)=-\Delta((L-aE)^2+{\cal C})+(E(r^2+a^2)-La)^2 \ \ and \ \ T(\theta)={\cal C}+cos^2\theta(E^2a^2-{{1}\over{sin^2\theta}}L^2).
\end{equation}

\

For {\it equatorial null geodesics} ({\it e.n.g.}), $\theta(\lambda)=\pi/2$ for all $\lambda$, so $\Sigma=r^2$, $\dot{\theta}(\lambda)=0$, and $T(\theta)={\cal C}$. From $p_\theta=g_{\theta\nu}\dot{x}^\nu=g_{\theta\theta}\dot{\theta}=-\Sigma\dot{\theta}$, $p_\theta^2=\Sigma^2\dot{\theta}^2=0=({{\partial S}\over{\partial\theta}})^2=({{\partial \tilde{S}_2(\theta)}\over{\partial\theta}})^2=T(\theta)$, it turns out that the Carter constant vanishes i.e. ${\cal C}=0$, and one has
\begin{equation}
R(r)=-\Delta(L-aE)^2+(E(r^2+a^2)-La)^2.
\end{equation}
The equation for $\dot{r}$ is obtained from $p_r=g_{r\nu}\dot{x}^\nu=g_{rr}\dot{r}=-{{\Sigma}\over{\Delta}}\dot{r}={{\partial \tilde{S}_1(r)}\over{\partial r}}$; so, from (16),
\begin{equation}
\dot{r}^2={{R(r)}\over{\Sigma^2}}={{R(r)}\over{r^4}}.
\end{equation}

\

The {\it principal equatorial null geodesics} {\it p.e.n.g.} are defined by the condition $L=aE$ i.e.
\begin{equation}
{{L}\over{E}}={{J}\over{M}}.
\end{equation}
That is, are the equatorial null geodesics in which the total angular momentum/unit of energy equals the black hole angular momentum/unit of its mass. In this case
\begin{equation}
\dot{r}=\pm E
\end{equation} 
and so 
\begin{equation}
\dot{t}={{r^2+a^2}\over{\Delta}}E \ \ and \ \ \dot{\psi}={{a}\over{\Delta}}E.
\end{equation}  
Therefore, {\it p.e.n.g.} are characterized by the tangent vectors 
\begin{equation}
k^\mu=(\dot{t},\dot{r},\dot{\theta},\dot{\psi})=({{r^2+a^2}\over{\Delta}}E,\pm E,0,{{a}\over{\Delta}}E).
\end{equation}
Rescaling the affine parameter $\lambda$ through the affine transformation $\lambda=\bar{\lambda}/E$, we obtain
\begin{equation}
k^\mu=(\dot{t},\dot{r},\dot{\theta},\dot{\psi})=({{r^2+a^2}\over{\Delta}},\pm 1,0,{{a}\over{\Delta}}).
\end{equation}
(The new affine parameter $\bar{\lambda}$ is again denoted by $\lambda$.) The associated 1-forms are $k_\mu=(-1,\pm{{r^2}\over{\Delta}},0,a)$. 

\

The + sign designs {\it outgoing principal equatorial null geodesics} ({\it o.p.e.n.g.}), with affine parameter $\lambda=r$, while the - sign designs {\it ingoing principal equatorial null geodesics} ({\it i.p.e.n.g.}), with affine parameter $\lambda=-r$.\

\

From ${{dt}\over{d\lambda}}={{\dot{t}}\over{\dot{r}}}=\pm {{r^2+a^2}\over{\Delta}} \ \ and \ \ {{d\psi}\over{dr}}={{\dot{\psi}}\over{\dot{r}}}=\pm{{a}\over{\Delta}}$, one has $dt=\pm{{r^2+a^2}\over{\Delta}}dr$, $d\psi=\pm {{a}\over{\Delta}}dr$ which imply 
\begin{equation}
t(r)=\pm\int^rdr^\prime{{{r^\prime}^2+a^2}\over{\Delta(r^\prime)}}+const., \ \ \psi(r)=\pm a\int^r{{dr^\prime}\over{\Delta(r^\prime)}}+const.^\prime
\end{equation}
Using (2.172) and (2.175.4) in [12], we finally obtain				
\begin{equation}
t(r)=\pm(r+{{Mr_+}\over{\sqrt{M^2-a^2}}}ln\vert {{r-r_+}\over{2M}}\vert-{{Mr_-}\over{\sqrt{M^2-a^2}}}ln\vert {{r-r_-}\over{2M}}\vert)+const.
\end{equation}
and 
\begin{equation}
\psi(r)=\pm {{a}\over{2\sqrt{M^2-a^2}}}ln\vert{{r-r_+}\over{r-r-}}\vert+const.^\prime 
\end{equation}
The plot of $t=t(r)$ in the $(t,r)$ plane, with $t,r\in(-\infty,+\infty)$, $r=0$ ($t$ axis) representing the {\it ring singularity} (which does not belong to spacetime), shows that (Fig. \ref{fig:1}):

\begin{figure}[H]
	\centering
	\includegraphics[width=0.8\linewidth]{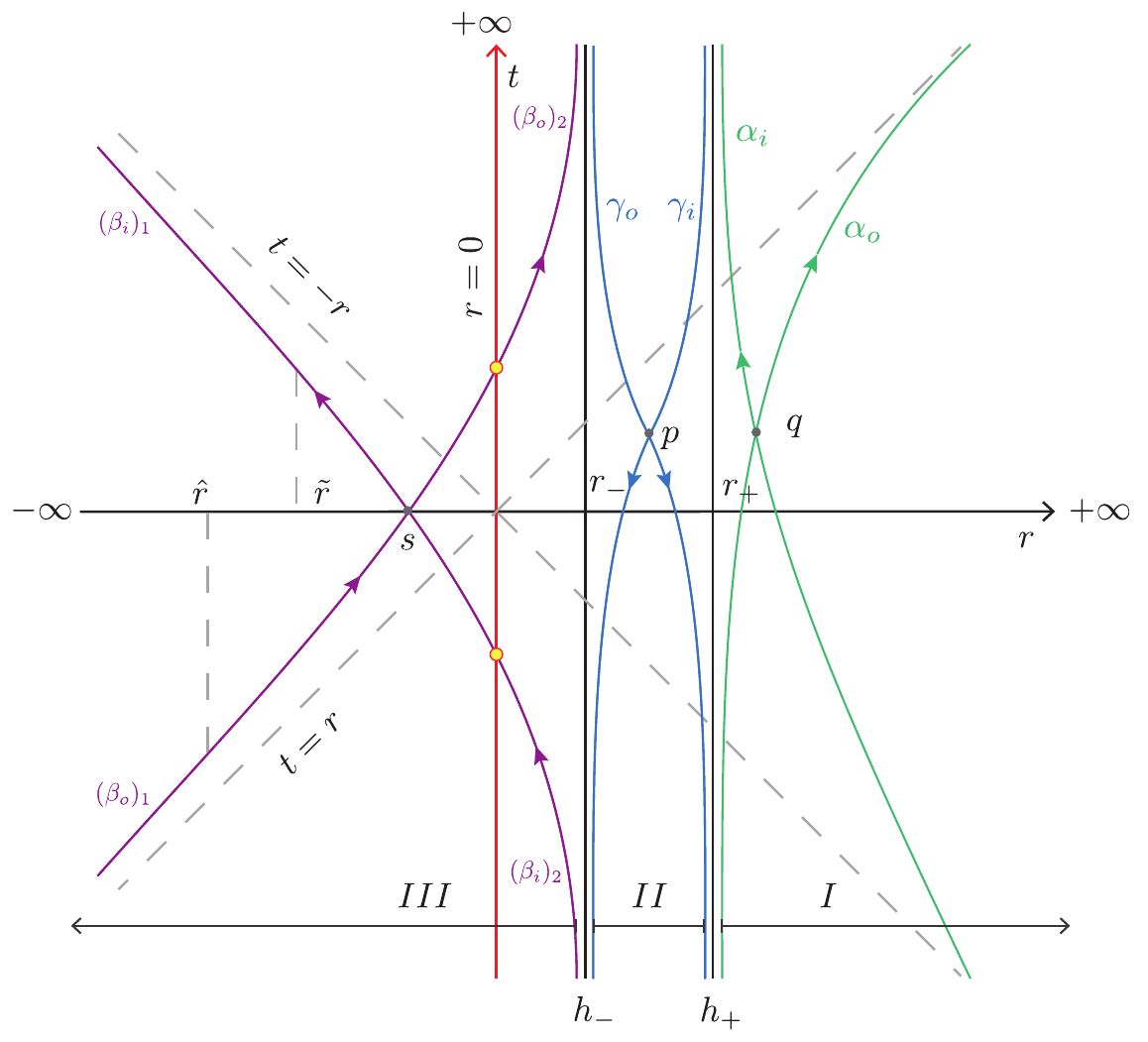} 
	\caption{ Principal equatorial null geodesics (\textit{p.e.n.g.}'s) in the $t/r$ plane}
	\label{fig:1}
\end{figure}

i) For each point $q$ in the region I: $r_+<r$, pass two and only two {\it forward} directed (one outgoing, $\alpha_o$, the other ingoing, $\alpha_i$) principal equatorial null geodesics ({\it f.d.o./i.p.e.n.g.}) (no past directed geodesics). Both $\alpha_o$ and $\alpha_i$ are inextendible i.e. future and past inextendible.

\

ii) For each point $p$ in the region II: $r_-<r<r_+$, pass two and only two {\it past} directed (one outgoing, $\gamma_o$, the other ingoing, $\gamma_i$) principal equatorial null geodesics ({\it p.d.o/i.p.e.n.g.}) (no future directed geodesics). Both $\gamma_o$ and $\gamma_i$ are inextendible. Clearly, neither $\gamma_o$ nor $ \gamma_i$ can be considered physical geodesics since they  propagate backwards in time.

\

iii) For each point $s$ in the region III: $-\infty<r<r_-$, pass two and only two {\it future} directed (one outgoing, $\beta_o$), the other ingoing, $\beta_i$) principal equatorial null geodesics ({\it f.d.o./i.p.e.n.g.}) (no past directed geodesics). Both geodesics of each pair split at the singularity: $\beta_o=(\beta_o)_1\cup(\beta_o)_2$, $\beta_i=(\beta_i)_1\cup(\beta_i)_2$ with $(\beta_o)_1\cap(\beta_o)_2=(\beta_i)_1\cap(\beta_i)_2=\phi$. $(\beta_o)_1$, $(\beta_i)_2$, $(\beta_o)_2$, and $(\beta_i)_1$ are inextendible.

\

All these geodesics are asymptotic to an horizon: $\alpha_i$, $\alpha_o$, and $\gamma_i$ to $h_+$; $\gamma_0$, $(\beta_o)_2$, and $(\beta_i)_2$ to $h_-$. It is interesting to note that the region where there are no {\it f.d.p.e.n.g.} is the unique region where trapped surfaces exist: $r\in(r_-,r_+)$. In a Kruskal-Szekeres extension, this region corresponds to the black and white holes.

\

\begin{figure}[H]
	\centering
	\includegraphics[width=0.8\linewidth]{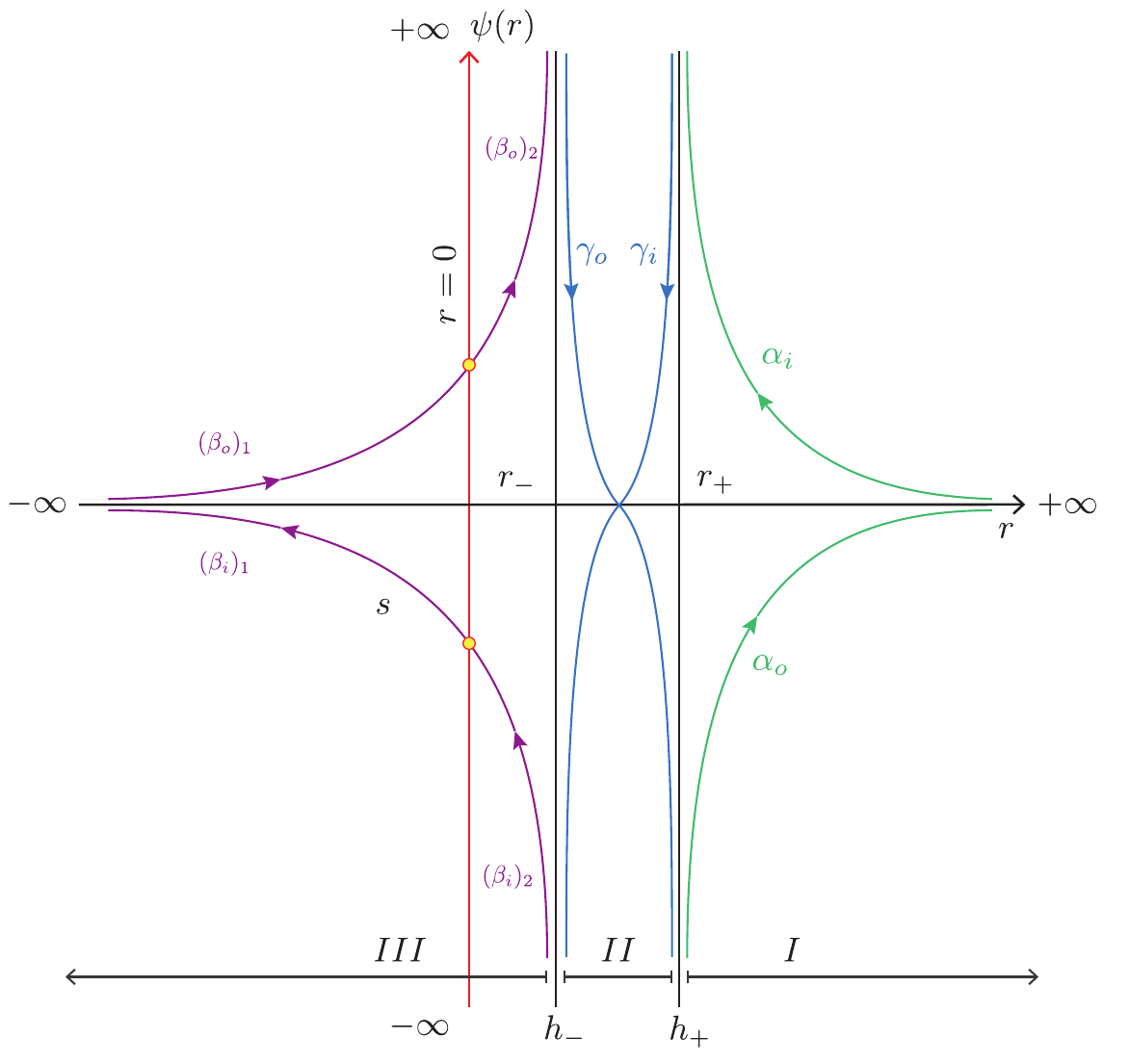} 
	\caption{Azimuthal angle as a function of $r$ of {\it p.e.n.g.}'s}
	\label{fig:2}
\end{figure}

The behavior of the azimuthal angle $\psi$ (eq.(28)) for each pair of the geodesics discussed above is shown in Fig. \ref{fig:2}. We notice that for forward directed geodesics $\dot{\psi}(r)\to -\infty$ as $r\to (r_+)_+$ and $r\to (r_-)_-$, and $\dot{\psi}(r)\to 0_-$ as $r\to \pm\infty$, in the ingoing cases, while $\dot{\psi}(r)\to +\infty$ as $r\to (r_+)_+$ and $r\to (r_-)_-$ and $\dot{\psi}(r)\to 0_+$ as $r\to \pm\infty$, in the outgoing cases. For past directed geodesics, $\dot{\psi}(r)\to +\infty$ as $r\to (r_+)_-$ and 
$r\to (r_-)_+$ in the ingoing case, and $\dot{\psi}(r)\to -\infty$ as $r\to (r_-)_+$ and $r\to (r_+)_-$ in the outgoing case.

\section{Congruences, expansions, and Raychaudhuri equations}

\

Varying the constants in (27) and (28) allows us to define geodesic congruences, which are governed by the Raychaudhuri equation [2]
\begin{equation}
{{d\Theta}\over{d\lambda}}=-{{1}\over{2}}\Theta^2-\sigma_{\mu\nu}\sigma^{\mu\nu}+\omega_{\mu\nu}\omega^{\mu\nu}-R_{\mu\nu}\dot{x}^\mu\dot{x}^\nu,
\end{equation} 
where 
\begin{equation}
\Theta=k^\mu_{;\mu}={{1}\over{\sqrt{-g}}}\partial_\mu(\sqrt{-g}k^\mu)
\end{equation}
is the expansion scalar ($g=det(g_{\mu\nu}))$; $\sigma_{\mu\nu}$, the shear, is the traceless symmetric part of the tensor $B_{\mu\nu}=D_\nu k_\mu$; $\omega_{\mu\nu}$, the rotation, is the antisymmetric part of $B_{\mu\nu}$; and $R_{\mu\nu}$ is the Ricci tensor. In vacuum, and in the absence of a cosmological constant, $R_{\mu\nu}=0$. Since the Kerr metric is an algebraically special solution in the Petrov's classification scheme [6], the Goldberg-Sachs theorem [7] implies that $\sigma_{\mu\nu}=0$. A direct calculation of $B_{\mu\nu}=k_{\mu,\nu}-\Gamma_{\mu\nu}^\rho k_\rho$ for {\it p.e.n.g.'s} gives $B_{\mu\nu}=B_{\nu\mu}$, which implies $\omega_{\mu\nu}=0$. A straightforward calculation of $\Theta$ gives 
\begin{equation}
\Theta_\pm=\pm{{2}\over{r}}
\end{equation}
with the upper (lower) sign for outgoing, $\lambda=r$ (ingoing, $\lambda=-r$) geodesics. For (29) one has
\begin{equation}
{{d\Theta}\over{d\lambda}}+{{1}\over{2}}\Theta^2=0.
\end{equation}

 \
 
 For future and past directed, outgoing and ingoing geodesic congruences defined by $\alpha$'s, $\beta$'s and $\gamma$'s, $\Theta(r)$ is finite for all $r>0$. Instead, for:
 
 \
  
  $(\beta_o)_2$: $r=0_+\to(r_-)_-$, the {\it f.d.o.p.n.g.}'s birth at the singularity at $r=0_+$, $\theta=\pi/2$ in the region III, and go asymptotically towards $(r_-)_-$; $\Theta_+\to+\infty$ as $r\to 0_+$;
  
  \
  
  $(\beta_i)_1$: $r=0_-\to r=-\infty$, the {\it f.d.i.p.n.g.}'s birth at the singularity at $r=0_-$, $\theta=\pi/2$ in the region III, and go asymptotically towards $r=-\infty$ along the line $t=-r$; $\Theta_-\to+\infty$ as $r\to 0_-$;
   
  \
   
  $(\beta_i)_2$: $(r_-)_-\to r=0_+$, the {\it f.d.i.p.n.g.}'s reach the singularity at $r=0_+$, $\theta=\pi/2$ in the region III, coming asymptotically from $(r_-)_-$; $\Theta_-\to-\infty$ as $r\to 0_+$;
    
  \
    
  $(\beta_o)_1$: $r=-\infty\to r=0_-$, the {\it f.d.o.p.n.g.}'s reach the singularity at $r=0_-$, $\theta=\pi/2$ in the region III, coming asymptotically along the line $t=r$ from $r=-\infty$; $\Theta_+\to-\infty$ as $r\to 0_+-$.
  
  \
  
  As expected, the only place where the expansions diverge is at the singularity ring.
  
  \
  
  Defining the functions $F_\pm(\lambda)$ through [3]
  \begin{equation}\Theta_\pm(\lambda)=2{{\dot{F}_\pm(\lambda)}\over{F_\pm(\lambda)}},
  \end{equation}
  (32) becomes the equation of a free non relativistic particle
  \begin{equation}
  \ddot{F}=0
  \end{equation}
with solution $F(\lambda)=v\lambda +w$, with $v$ and $w$ real constants. For the cases of interest $(\beta_i)_s$, $(\beta_o)_s$, $s=1,2$, $\Theta(0)=\pm\infty$ ($0=0_+$ or $0_-$) implies $F(0)=0$ if $\dot{F}(0)\neq 0$. Then $w=0$ and $F(\lambda)=v\lambda$, $v\neq 0$. (Essentially, $F$ is a measure of the transverse area of the congruence.)

\

The Lagrangian that reproduces (34) is ${{1}\over{2}}(\dot{F}(\lambda))^2$, with an associated action 
\begin{equation}
S[F]={{1}\over{2}}\int_{\lambda '}^{\lambda  }(\dot{F}(\lambda))^2={{v^2}\over{2}}(\lambda ''-\lambda ').
\end{equation}

(Units: $[F]=[length]^{1/2}$ since $[S]=[length]^0$; then $[v]=[length]^{-1/2}$.)

\

\section{Quantum propagators}

\

For each geodesic congruence associated to the geodesics $\alpha_i$, $\alpha_o$, $\gamma_i$, $\gamma_o$, $(\beta_i)_2$, $(\beta_i)_1$, $(\beta_o)_2$, and $(\beta_o)_1$, one can formally associate the ``quantum" (Feynman) propagator [13]
\begin{equation}
K(F'',\lambda'';F',\lambda')={{1}\over{\sqrt{2\pi i(\lambda''-\lambda')}}}e^{{{i}\over{2}}v^2(\lambda''-\lambda')}.
\end{equation}
The only propagators where divergences might appear as $r\to 0$ ($0=0_+$ or $0_-$) are those associated to $(\beta_i)_s$, $(\beta_o)_s$, $s=1,2$. We study these cases in detail.

\

$(\beta_i)_2$: $F''=v_-\lambda''$, $\lambda''=0_+$; $F'=v_-\lambda'$, $\lambda'=(r_-)_-\equiv r_-$; then
\begin{equation}
(K_i)_2(0,0_+;v_-r_-,r_-)={{e^{-iv_-^2r_-/2}}\over{\sqrt{-2\pi ir_-}}}={{e^{-i(v_-^2r_--\pi/2)/2}}\over{\sqrt{2\pi r_-}}}, \ \ with \ \ \vert(K_i)_2\vert=1/\sqrt{2\pi r_-}<+\infty.
\end{equation}

\

$(\beta_i)_1$: $F''=v_-\tilde{r}$, $\lambda''=\tilde{r}$; $F'=v_-\lambda'$, $\lambda'=0_-$; then
\begin{equation}
(K_i)_1(v_-\tilde{r},0_+;0,0_-)={{e^{iv_-^2\tilde{r}/2}}\over{\sqrt{2\pi i\tilde{r}}}}={{e^{i(v_-^2\tilde{r}-\pi/2)/2}}\over{\sqrt{2\pi \tilde{r}}}}, \ \ with \ \ \vert(K_i)_1\vert=1/\sqrt{2\pi \tilde{r}}<+\infty.
\end{equation} 

\

$(\beta_o)_2$: $F''=v_+\lambda''$, $\lambda''=(r_-)_-\equiv r_-$; $F'=v_+\lambda'$, $\lambda'=0_+$; then
\begin{equation}
(K_o)_2(v_+r_-,r_-;0,0_+)={{e^{iv_+^2r_-/2}}\over{\sqrt{2\pi ir_-}}}={{e^{i(v_+^2r_--\pi/2)/2}}\over{\sqrt{2\pi r_-}}}, \ \ with \ \ \vert(K_o)_2\vert=1/\sqrt{2\pi r_-}<+\infty.
\end{equation}

\

$(\beta_o)_1$: $F''=v_+\lambda''$, $\lambda''=0_-$; $F'=\hat{r}v_+$, $\lambda'=\hat{r}$; then
\begin{equation}
(K_o)_1(0,0_-;\hat{r}v_+,\hat{r})={{e^{-iv_+^2\hat{r}/2}}\over{\sqrt{-2\pi i\hat{r}}}}={{e^{-i(v_+^2\hat{r}-\pi/2)/2}}\over{\sqrt{2\pi \hat{r}}}}, \ \ with \ \ \vert(K_o)_1\vert=1/\sqrt{2\pi \hat{r}}<+\infty.
\end{equation}

\

The $\pm$ signs in $v_\pm$ indicate outgoing and ingoing geodesics; $v_\pm\neq 0$ are arbitrary real constants; and in the $\tilde{r},\hat{r}\to -\infty$ limit the corresponding propagators vanish since the infinite oscillations of the exponentials are ``killed" by the growing of the denominators. The similarity between the propagators $(K_i)_2$ and $(K_o)_1$ (($K_i)_1$ and $K_o)_2$) is due to the fact that the congruences defined by $(\beta_i)_2$ and $(\beta_o)_1$ ( $(\beta_i)_1$ and $(\beta_o)_2$) arrive (birth) at the singularity.

\

The results (37)-(40) prove that, even if the expansion scalars $\Theta$ which govern the classical evolution of the geodesic congruences diverge at the singularity ring, the associated Feynman propagators remain finite, which is an indication that at a quantum level singularities might disappear or at least become softened.

\

\section{Conclusion and final remarks}

\

We showed that to any principal equatorial null geodesic congruence in Kerr spacetime, can be asigned a quantum (Feynman) propagator describing its flow, which is classically described by the Raychaudhuri equation for the expansion scalar $\Theta$. In particular, the unique potentially divergent propagators, those reaching the ring singularity at $(r=0,\theta=\pi/2)$, i.e. those in the region $-\infty<r<r_-$, remain finite, in contradistinction with the expansion scalars which diverge as $1/r$. This fact suggests that at the quantum level, singularities which appear at the classical level, might not be present or, at least, might be smoothened. The same conclusion is arrived by S. Chakraborty and M. Chakraborty in their recent review [16].

\

As final remarks we want to mention two facts. First, the only part of the Kerr solution which should represent the result of the collapse of a rotating star, that is, the only physical part, is the previously called ``shell horizon" ($r_-<r<r_+$) which in particular also is the unique globally hyperbolic region. In a Penrose diagram, it looks as a central ``diamond" [14]. Paradoxically, it is free of singularities, even if it contains the black and white holes! Moreover, in the asymptotically flat ($r\to-\infty$) regions beyond $r_-$ and close to the singularity rings ($r<0$, $r\sim 0$), closed timelike curves exist that violate causality [15]. This criterium for defining the physical region should make the divergence of $\Theta$ as $r\to0_-$ totally harmless. Second, a spacetime is considered singular, if, inextendible, it has at least one causal (in particular null) inextendible geodesic which does not admit an affine parameter extending from $-\infty$ to $+\infty$. This is precisely the case of the inextendible geodesics $\gamma_o$ and $\gamma_i$ with affine parameter $r\in(r_-,r_+)$. Any other affine parameter must be of the form $\lambda=xr+y$ with $x,y\in \mathbb{R}$, $x\neq 0$, which is always finite. Since both $\gamma_o$ and $\gamma_i$ are non physical, the physical spacetime (the diamond) can still be considered non singular. 

\

{\bf Acknowledgments}

\

One of us (M.S.) thanks for hospitality to the Instituto de Astronom\'\i a y F\'\i sica del Espacio (IAFE) de la Universidad de Buenos Aires and CONICET, Argentina, where this work was done during a sabbatical stay. The authors thank Oscar Brauer at the University of Leeds, U.K., for the drawing of Figures 1 and 2.

\

{\bf Conflicts of interest}

\

The authors declare no conflicts of interest regarding the publication of this paper.

\

{\bf References}

\

[1] Socolovsky, M. (2021) Quantum Propagators of Geodesic Congruences. {\it Theoretical Physics}, {\bf 6}, No. 2, 9-17. 

https://doi.org/10.22606/tp.2021.62001; arXiv: 2108.03115v1 [gr-qc], https://doi.org/1048550/arXiv.2108.03115

\

[2] Raychaudhuri, A. (1955) Relativistic Cosmology I. {\it Physical Review}, {\bf 98}, 1123-1126. 

https://doi.org/10.1103/PhysRev.98.1123

\

[3] Tipler, F.J. (1978) General Relativity and Conjugate Ordinary Differential Equations. {\it Journal of Differential Equations}, {\bf 30}, 165-174. https://doi.org/10.1016/0022-0396(78)90012-8

\

[4] \'Alvarez, E. (2021) Windows on Quantum Gravity. {\it Fortschritte der Physik}, {\bf 69}, 2000080, 1-29.

https://doi.org/10.1002/prop.202000080

\

[5] Kerr, R.P. (1963) Gravitational field of a spinning mass as an example of algebraically special metrics. {\it Physical Review Letters}, {\bf 11}, 237-238. https://doi.org/10.1103/PhysRevLett.11.237

\

[6] Petrov, A.Z. (1954) Classification of spaces defined by gravitational fields. English translation: (2000), {\it General Relativity and Gravitation}, {\bf 32}, 1665-1685. https://doi.org/10.1023/A:1001910908054

\

[7] Goldberg, J.N. and Sachs, R.K. (1962) A theorem of Petrov types. Republished: (2009), {\it General Relativity and Gravitation}, {\bf 41}, 433-444. https://doi.org/10.1007/s10714-008-0722-5

\

[8] Ferrari, V., Gualtieri, L. and Pani, P. (2020) General Relativity and its Applications. Black Holes, Compact Stars and Gravitational Waves. CRC Press.

\

[9] Carter, B. (1968) Global structure of the Kerr family of gravitational fields. {\it Physical Review}, {\bf 174}, 1559-1571. 

https://doi.org/10.1103/PhysRev.174.1559

\

[10] Kerr, R.P. (2023) Do black holes have singularities? {\it arXiv}: [{\it gr-qc}] 2312.00841v1. https://doi.org/10.48550/arXiv.2312.00841

\

[11] Krtous, P., Frolov, V.P., and Kubiznák, D. (2008) Hidden symmetries of higher dimensional black holes and uniqueness of the Kerr-NUT-(A)dS spacetime. {\it Physical Review D}, {\bf 78}, 064022, 1-5. https://doi.org/10.1103/PhysRevD.78.064.022

\

[12] Gradshtein, I.S. and Ryzhik, I.M. (1980) Table of Integrals, Series, and Products, Academic Press, San Diego.

\

[13] Feynmann, R.P. and Hibbs, A.R. (1965) Quantum Mechanics and Path Integrals, McGraw-Hill, N.Y. 

\

[14] Witten, E. (2020) Light Rays, Singularities, and All That. {\it Reviews of Modern Physics}, {\bf 92}, 045004. 

\
https://doi.org/10.1103/RevModPhys.92.045004

\

[15] Thorne, K.S. (1993) Closed Timelike Curves. {\it Proceedings of the 13th International Conference on General Relativity and Gravitation}, ed. C. Kozameh. CRC Press.

\

[16] Chakraborty, S. and Chakraborty, M. (2024) A Revisit to Classical and Quantum aspects of Raychaudhuri equation and possible solutions of Singularity. arXiv: 2402.17799v1 [gr-qc]

\end{document}